\documentclass[12pt,a4paper]{article}
\begin{document}

\baselineskip 18pt
\newcommand{\sheptitle}
{Solving the BFKL Equation in the Next--to--Leading Approximation}

\newcommand{\shepauthor}
{Jeppe R.~Andersen$^{\rm a,b}$, Agust{\'\i}n Sabio Vera$^{\rm a}$}

\newcommand{\shepaddress}
{$^{\rm a}$Cavendish Laboratory,
  University of Cambridge, Madingley Road, CB3 0HE,Cambridge, UK\\
$^{\rm b}$DAMTP, Centre for Mathematical Sciences,
  Wilberforce Road, CB3 0WA,Cambridge, UK}

\newcommand{\shepabstract} {The Balitsky--Fadin--Kuraev--Lipatov equation in
  the next--to--leading logarithmic approximation is solved using an
  iterative method. We derive the solution for forward scattering with all
  conformal spins. A discussion of the infrared finiteness of the results is
  included.}

\begin{titlepage}
\begin{flushright}
Cavendish-HEP-2003/04\\
DAMTP-2003-46\\
\end{flushright}
\begin{center}
{\large{\bf \sheptitle}}
\bigskip \\ \shepauthor \\ \mbox{} \\ {\it \shepaddress} \\ \vspace{.5in}
{\bf Abstract} \bigskip \end{center} \setcounter{page}{0}
\shepabstract
\end{titlepage}

\section{Introduction}

The Balitsky--Fadin--Kuraev--Lipatov (BFKL) \cite{FKL} formalism enables the
resummation of logarithms appearing in scattering amplitudes, which are large
in the Regge limit, where the center of mass energy $\sqrt{s}$ is large and
the momentum transfer $\sqrt{-t}$ fixed.  In this approach the high energy
cross-section for the process $A+B \rightarrow A'+B'$ is factorised as
\begin{eqnarray}
\label{cross--section1}
\sigma(s) &=&\int 
\frac{d^2 {\bf k}_a}{2 \pi{\bf k}_a^2}
\int \frac{d^2 {\bf k}_b}{2 \pi {\bf k}_b^2} ~\Phi_A({\bf k}_a) ~\Phi_B({\bf k}_b)
~f \left({\bf k}_a,{\bf k}_b, \Delta = \ln{\frac{s}{s_0}}\right),
\end{eqnarray}
where $\Phi_{A,B}$ are process--dependent impact factors and $f\left({\bf
    k}_a,{\bf k}_b,\Delta\right)$ is the process--independent gluon Green's
function describing the interaction between two Reggeised gluons exchanged in
the $t$--channel with transverse momenta ${\bf k}_{a,b}$. In this Letter we
use the Regge scale $s_0 = \left|{\bf k}_a\right|\left|{\bf k}_b\right|$, a
different choice would modify the impact factors in such a way that the
prediction for the cross--section remains unchanged.

The representation in (\ref{cross--section1}) is valid \cite{LF} in the
leading logarithmic approximation (LLA), where terms of the form
$\left(\alpha_s \Delta \right)^n$ are resummed, and in the next--to--leading
logarithmic approximation (NLLA) \cite{Fadin:1998py}, where contributions of
the type $\alpha_s \left(\alpha_s\Delta \right)^n$ are also taken into
account. This formalism is valid in both the forward and non--forward cases
\cite{nonforward}. In this Letter we deal with the former but the proposed
method is also applicable to the latter.

In recent years there have been many studies of the behaviour of the gluon
Green's function in the NLLA \cite{NLLpapers}. This Green's function is
obtained as the solution of an integral equation where radiative corrections
enter through its kernel. The main difficulty in solving the equation
analytically in the NLLA stems from the logarithmic dependence introduced by
the running of the coupling. In this work we show how it is possible to solve
the equation using an iterative method directly in energy space, considering
the full kernel with scale invariant and running coupling terms. In this
approach we keep all the angular information from the BFKL evolution, solving
the equation for a general conformal spin without relying on any asymptotic
expansion.

In Section \ref{BFKL@NLLA} we present the BFKL equation in the NLLA in
dimensional regularisation and show how, introducing a cut--off in the real
emissions, it is possible to write it in a form suitable for iteration. In
Section \ref{Iterating} we solve the equation using an iterative method and
compare with previous results in the literature. In Section \ref{conclusions}
we present our conclusions.

\section{The BFKL equation in the NLLA}
\label{BFKL@NLLA}

To write the BFKL equation for the gluon Green's function, $f \left({\bf
    k}_a,{\bf k}_b, \Delta\right)$, it is convenient to introduce the Mellin
transform, $f_\omega \left({\bf k}_a,{\bf k}_b\right)$, in $\Delta$ space as
follows
\begin{eqnarray}
\label{Mellin}
f \left({\bf k}_a,{\bf k}_b, \Delta\right) 
&=& \frac{1}{2 \pi i}
\int_{a-i \infty}^{a+i \infty} d\omega ~ e^{\omega \Delta} f_{\omega} 
\left({\bf k}_a ,{\bf k}_b\right).
\end{eqnarray}
In this way the BFKL equation in the NLLA for forward scattering can be
written in dimensional regularisation $\left(D = 4+2 \epsilon \right)$ as
\begin{eqnarray}
\omega f_\omega \left({\bf k}_a,{\bf k}_b\right) &=& \delta^{(2+2\epsilon)} 
\left({\bf k}_a-{\bf k}_b\right) + \int d^{2+2\epsilon}{\bf k}' ~
\mathcal{K}\left({\bf k}_a,{\bf k}'\right)f_\omega \left({\bf k}',{\bf k}_b 
\right).
\end{eqnarray}

The kernel in the NLLA is expressed in terms of the gluon Regge trajectory
\cite{trajectory}, $\omega^{(\epsilon)} \left({\bf k}_a^2\right)$, and the
real emission component \cite{emission}, $\mathcal{K}_r \left({\bf k}_a,{\bf
    k}\right)$, in such a way that
\begin{eqnarray}
\mathcal{K}\left({\bf k}_a,{\bf k}\right) = 2 \,\omega^{(\epsilon)}\left({\bf k}_a^2\right) \,\delta^{(2+2\epsilon)}\left({\bf k}_a-{\bf k}\right) + \mathcal{K}_r\left({\bf k}_a,{\bf k}\right).
\end{eqnarray}
It is convenient to split the kernel $\mathcal{K}_r$ into two parts: a
$\epsilon$--dependent, $\mathcal{K}_r^{(\epsilon)}$, and a
$\epsilon$--independent, $\widetilde{\mathcal{K}}_r$. 
In the integral over real emission we also perform a shift
in the variable of integration over transverse degrees of
freedom of the form ${\bf k} = {\bf k}' - {\bf k}_a$ to
write
\begin{eqnarray}
\omega f_\omega \left({\bf k}_a,{\bf k}_b\right) &=& \delta^{(2+2\epsilon)}
\left({\bf k}_a-{\bf k}_b\right) + \int d^{2+2\epsilon}{\bf k} \, 2 \,
\omega^{(\epsilon)} \left({\bf k}_a^2\right) \delta^{(2+2\epsilon)} 
\left({\bf k}_a-{\bf k}\right) f_\omega \left({\bf k},{\bf k}_b\right) \\
&&\hspace{-3.2cm}+ \int d^{2+2\epsilon}{\bf k} \, \mathcal{K}_r^{(\epsilon)} \left({\bf k}_a,{\bf k}_a+{\bf k}\right) f_\omega \left({\bf k}_a+{\bf k},{\bf k}_b\right)+ 
\int d^{2+2\epsilon}{\bf k} \, \widetilde{\mathcal{K}}_r \left({\bf k}_a,{\bf k}_a+{\bf k}\right) f_\omega \left({\bf k}_a+{\bf k},{\bf k}_b\right).\nonumber
\end{eqnarray}

In order to explicitly show the cancellation of the poles at $\epsilon = 0$
we split the integral over transverse phase space for
$\mathcal{K}_r^{(\epsilon)}$ into two regions separated by a cut--off
$\lambda$, i.e.
\begin{eqnarray}
\label{cut--off}
\omega f_\omega \left({\bf k}_a,{\bf k}_b\right) &=& \delta^{(2+2\epsilon)}
\left({\bf k}_a-{\bf k}_b\right) + \int d^{2+2\epsilon}{\bf k} \, 2 \,
\omega^{(\epsilon)} \left({\bf k}_a^2\right) \delta^{(2+2\epsilon)} 
\left({\bf k}_a-{\bf k}\right) f_\omega \left({\bf k},{\bf k}_b\right) \nonumber\\
&&\hspace{-1cm}+ \int d^{2+2\epsilon}{\bf k} \, \mathcal{K}_r^{(\epsilon)} \left({\bf k}_a,{\bf k}_a+{\bf k}\right) \left(\theta\left({\bf k}^2-\lambda^2\right)+\theta\left(\lambda^2-{\bf k}^2 \right)\right)f_\omega \left({\bf k}_a+{\bf k},{\bf k}_b\right)\nonumber\\
&&\hspace{-1cm} + \int d^{2+2\epsilon}{\bf k} \, \widetilde{\mathcal{K}}_r \left({\bf k}_a,{\bf k}_a+{\bf k}\right) f_\omega \left({\bf k}_a+{\bf k},{\bf k}_b\right).
\end{eqnarray}
For any finite $\lambda$ we now introduce an arbitrary dependence of
$\mathcal{O}\left(\frac{\lambda^2}{{\bf k}_a^2}\right)$ by using the
approximation $f_\omega \left({\bf k}_a+{\bf k},{\bf k}_b\right) \simeq
f_\omega \left({\bf k}_a,{\bf k}_b\right)$ for $\left|{\bf k}\right| <
\lambda$. This dependence is negligible when the external scale ${\bf k}_a^2$
is large. It is possible to introduce the cut--off in several different ways
but all the choices must have the same $\lambda \rightarrow 0$ limit. In this
way we can approximate Eq.~(\ref{cut--off}) by
\begin{eqnarray}
\omega f_\omega \left({\bf k}_a,{\bf k}_b\right) &=& \delta^{(2+2\epsilon)}
\left({\bf k}_a-{\bf k}_b\right) \\
&&\hspace{-2cm}+ \left\{2 \, 
\omega^{(\epsilon)} \left({\bf k}_a^2\right) + \int d^{2+2\epsilon}{\bf k} \, \mathcal{K}_r^{(\epsilon)} \left({\bf k}_a,{\bf k}_a+{\bf k}\right) \theta\left(\lambda^2-{\bf k}^2\right) \right\}f_\omega \left({\bf k}_a,{\bf k}_b\right)\nonumber\\
&&\hspace{-2cm}+ \int d^{2+2\epsilon}{\bf k} \, \left\{ \mathcal{K}_r^{(\epsilon)} \left({\bf k}_a,{\bf k}_a+{\bf k}\right) \theta\left({\bf k}^2-\lambda^2\right) + \widetilde{\mathcal{K}}_r \left({\bf k}_a,{\bf k}_a+{\bf k}\right) \right\} f_\omega \left({\bf k}_a+{\bf k},{\bf k}_b\right).\nonumber
\end{eqnarray}

In dimensional regularisation the gluon Regge trajectory can be written as \cite{Fadin:1998py}
\begin{eqnarray}
2 \, \omega^{(\epsilon)} \left({\bf q}^2\right) &=& - \bar{\alpha}_s \frac{\Gamma (1-\epsilon)}{(4 \pi)^\epsilon} \left(\frac{1}{\epsilon}+\ln{\frac{q^2}{\mu^2}}\right) 
- \frac{\bar{\alpha}_s^2}{8}\frac{\Gamma^2(1-\epsilon)}{(4 \pi)^{2 \epsilon}}
\left\{\frac{\beta_0}{N_c}\left(\frac{1}{\epsilon^2}+\ln^2{\frac{q^2}{\mu^2}}\right)\right.\\
&+& \left.\left(\frac{4}{3}-\frac{\pi^2}{3}+\frac{5}{3}\frac{\beta_0}{N_c}\right)\left(\frac{1}{\epsilon}+2\ln{\frac{q^2}{\mu^2}}\right)-\frac{32}{9}+2 \zeta(3)-\frac{28}{9}\frac{\beta_0}{N_c}\right\},\nonumber
\end{eqnarray}
where $\beta_0 \equiv \frac{11}{3}N_c-\frac{2}{3}n_f$, $\bar{\alpha}_s \equiv
\frac{\alpha_s (\mu) N_c}{\pi}$ and $\alpha_s (\mu) = \frac{g_\mu^2}{4 \pi}$.
$\mu$ is the renormalisation scale in the $\overline{\rm MS}$ scheme.

The $\epsilon$-dependent part of the real emission kernel reads \cite{Fadin:1998py}
\begin{eqnarray}
\mathcal{K}_r^{(\epsilon)} \left({\bf q},{\bf q}+{\bf k}\right) &=&
\frac{\bar{\alpha}_s \mu^{-2 \epsilon}}{\pi^{1+\epsilon}(4 \pi)^\epsilon}
\frac{1}{{\bf k}^2} \left\{1+\frac{\bar{\alpha}_s}{4}\frac{\Gamma(1-\epsilon)}{(4 \pi)^\epsilon}\left[\frac{\beta_0}{N_c}\frac{1}{\epsilon}\left(1-\left(\frac{{\bf k}^2}{\mu^2}\right)^\epsilon \left(1-\epsilon^2 \frac{\pi^2}{6}\right) \right) \right.\right.\nonumber\\
&&\hspace{-1cm}+\left.\left.\left(\frac{{\bf k}^2}{\mu^2}\right)^\epsilon \left(\frac{4}{3}-\frac{\pi^2}{3}+\frac{5}{3}\frac{\beta_0}{N_c}+\epsilon\left(-\frac{32}{9}+14\zeta(3)-\frac{28}{9}\frac{\beta_0}{N_c}\right)\right)\right]\right\}.
\end{eqnarray}
For our purposes we are interested in the integration of this piece of the
kernel over the phase space limited by the cut--off, i.e.
\begin{eqnarray}
\int d^{2+2\epsilon}{\bf k} \, \mathcal{K}_r^{(\epsilon)} \left({\bf q},{\bf q}+{\bf k}\right) \theta\left(\lambda^2-{\bf k}^2\right) &=& \\
&&\hspace{-6.6cm}\frac{1}{\Gamma(1+\epsilon)}\frac{\bar{\alpha}_s}{(4 \pi)^\epsilon}
\frac{1}{\epsilon} \left(\frac{\lambda^2}{\mu^2}\right)^\epsilon\left\{1+\frac{\bar{\alpha}_s}{4}\frac{\Gamma(1-\epsilon)}{(4 \pi)^\epsilon}\left[\frac{\beta_0}{N_c}\frac{1}{\epsilon}\left(1-\frac{1}{2}\left(\frac{\lambda^2}{\mu^2}\right)^\epsilon \left(1-\epsilon^2 \frac{\pi^2}{6}\right) \right) \right.\right.\nonumber\\
&&\hspace{-6.6cm}+\left.\left.\frac{1}{2}\left(\frac{\lambda^2}{\mu^2}\right)^\epsilon \left(\frac{4}{3}-\frac{\pi^2}{3}+\frac{5}{3}\frac{\beta_0}{N_c}+\epsilon\left(-\frac{32}{9}+14\zeta(3)-\frac{28}{9}\frac{\beta_0}{N_c}\right)\right)\right]\right\}.\nonumber
\end{eqnarray}
When this term is combined with the trajectory, the poles in $\epsilon$
cancel and we obtain a finite expression depending on $\lambda$:
\begin{eqnarray}
\omega_0 \left({\bf q}^2,\lambda^2\right) &\equiv&\lim_{\epsilon \to 0} \left\{ 2\, \omega^{(\epsilon)}\left({\bf q}^2\right) + \int d^{2+2\epsilon}{\bf k} \,
\mathcal{K}_r^{(\epsilon)} \left({\bf q},{\bf q}+{\bf k}\right) 
\theta \left(\lambda^2-{\bf k}^2\right) \right\}~=~ \\
&&\hspace{-2cm}- \bar{\alpha}_s \left\{\ln{\frac{{\bf q}^2}{\lambda^2}}
+ \frac{\bar{\alpha}_s}{4}\left[\frac{\beta_0}{2 N_c}\ln{\frac{{\bf q}^2}{\lambda^2}}\ln{\frac{\mu^4}{{\bf q}^2 \lambda^2}}+\left(\frac{4}{3}-\frac{\pi^2}{3}+\frac{5}{3}\frac{\beta_0}{N_c}\right)\ln{\frac{{\bf q}^2}{\lambda^2}}-6 \zeta(3)\right]\right\}.\nonumber
\end{eqnarray}

To simplify our formulae we introduce the notation 
\begin{eqnarray}
\omega_0 \left({\bf q}^2,\lambda^2 \right) &\equiv& - \xi\left(\left|{\bf q}\right|\lambda\right) \ln{\frac{{\bf q}^2}{\lambda^2}} + \eta 
\end{eqnarray}
with 
\begin{eqnarray}
\xi \left({\rm X}\right) &\equiv& \bar{\alpha}_s +  
\frac{{\bar{\alpha}_s}^2}{4}\left[\frac{4}{3}-\frac{\pi^2}{3}+\frac{5}{3}\frac{\beta_0}{N_c}-\frac{\beta_0}{N_c}\ln{\frac{{\rm X}}{\mu^2}}\right] 
\end{eqnarray}
and
\begin{eqnarray}
\eta ~\equiv~ {\bar{\alpha}_s}^2 \frac{3}{2} \zeta (3). 
\end{eqnarray}
In this way we can write
\begin{eqnarray}
\lim_{\epsilon \to 0} \int d^{2+2\epsilon}{\bf k} \, \mathcal{K}_r^{(\epsilon)} \left({\bf k}_a,{\bf k}_a+{\bf k}\right) \theta\left({\bf k}^2-\lambda^2\right) f_\omega \left({\bf k}_a+{\bf k},{\bf k}_b\right) &=& \\
&&\hspace{-6cm} \int d^{2}{\bf k} \, \frac{1}{\pi {\bf k}^2} \xi \left({\bf k}^2\right) \theta\left({\bf k}^2-\lambda^2\right) f_\omega \left({\bf k}_a+{\bf k},{\bf k}_b \right).\nonumber
\end{eqnarray}

With these conventions, the forward BFKL equation in the NLLA can finally 
be written as
\begin{eqnarray}
\label{nll}
\left(\omega - \omega_0\left({\bf k}_a^2,\lambda^2\right)\right) f_\omega \left({\bf k}_a,{\bf k}_b\right) &=& \delta^{(2)} \left({\bf k}_a-{\bf k}_b\right)\\
&&\hspace{-4cm}+ \int d^2 {\bf k} \left(\frac{1}{\pi {\bf k}^2} \xi \left({\bf k}^2\right) \theta\left({\bf k}^2-\lambda^2\right)+\widetilde{\mathcal{K}}_r \left({\bf k}_a,{\bf k}_a+{\bf k}\right)\right)f_\omega \left({\bf k}_a+{\bf k},{\bf k}_b\right),\nonumber
\end{eqnarray}
where \cite{Fadin:1998py}
\begin{eqnarray}
\label{non_ang_av}
\widetilde{\mathcal{K}}_r \left({\bf q},{\bf q}'\right) &=& 
\frac{\bar{\alpha}_s^2}{4 \pi} 
\left\{-\frac{1}{({\bf q}-{\bf q'})^2}\ln^2{\frac{{\bf q}^2}{{\bf q'}^2}}+
\left(1+\frac{n_f}{N_c^3}\right)
\left(\frac{3({\bf q}\cdot{\bf q'})^2
-2 {\bf q}^2 {\bf q'}^2}{16 {\bf q}^2 {\bf q'}^2}\right) \right.\\
&&\hspace{6cm}\times \left(\frac{2}{{\bf q}^2}+\frac{2}{{\bf q'}^2}
+\left(\frac{1}{{\bf q'}^2}-\frac{1}{{\bf q}^2}\right)
\ln{\frac{{\bf q}^2}{{\bf q'}^2}}\right) \nonumber\\
&&\hspace{-2.2cm}-\left(3+\left(1+\frac{n_f}{N^3_c}\right)
\left(1-\frac{({\bf q}^2+{\bf q'}^2)^2}{8 {\bf q}^2 {\bf q'}^2}
- \frac{(2 {\bf q}^2 {\bf q'}^2 - 3 {\bf q}^4 - 3 {\bf q'}^4)}
{16 {\bf q}^4 {\bf q'}^4}({\bf q} \cdot {\bf q'})^2\right)\right)\nonumber\\
&&\hspace{6cm}\times 
\int^\infty_0 dx \frac{1}{{\bf q}^2 + x^2 {\bf q'}^2} \ln{\left|\frac{1+x}{1-x}\right|}\nonumber\\
&&\hspace{-2.2cm}
+\frac{2({\bf q}^2-{\bf q'}^2)}{({\bf q}-{\bf q'})^2({\bf q}+{\bf q'})^2} 
\left(\frac{1}{2}\ln{\frac{{\bf q}^2}{{\bf q'}^2}}
\ln{\frac{{\bf q}^2 {\bf q'}^2 ({\bf q}-{\bf q'})^4}
{({\bf q}^2+{\bf q'}^2)^4}}
+ \left( \int_0^{- \frac{{\bf q}^2}{{\bf q'}^2}} -
\int_0^{- \frac{{\bf q'}^2}{{\bf q}^2}} \right) 
dt \frac{\ln(1-t)}{t}\right)\nonumber\\
&&\hspace{-2.2cm}\left.-\left(1-\frac{({\bf q}^2-{\bf q'}^2)^2}{({\bf q}-{\bf q'})^2
({\bf q}+{\bf q'})^2}\right) 
\left( \left( \int_0^1 
-\int_1^\infty \right) dz \frac{1}{({\bf q'}-z {\bf q})^2}
\ln{\frac{(z {\bf q})^2}{{\bf q'}^2}}\right)\right\}.
\nonumber
\end{eqnarray}
It is important to note that this kernel contains the full angular
information in the BFKL evolution. Writing the equation as in (\ref{nll}) is
very natural in the sense that there is a clear separation between the
virtual contributions on the left hand side, and the real emissions,
integrated over phase space, on the right hand side.

To study the dependence on $\lambda$ of Eq.(16) we take the
derivative with respect to $\lambda^2$ of the
$\lambda$--dependent terms, i.e.
\begin{eqnarray}
\label{indep}
\frac{\partial}{\partial \lambda^2} \left\{\omega_0 \left({\bf k}_a^2,\lambda^2\right) f_\omega \left({\bf k}_a,{\bf k}_b\right) + \int d^2 {\bf k} \frac{1}{\pi {\bf k}^2}\xi \left({\bf k}^2\right) \theta \left({\bf k}^2-\lambda^2\right) f_\omega \left({\bf k}_a+{\bf k},{\bf k}_b\right)\right\} &=& \nonumber\\
\frac{1}{\lambda^2} \xi \left(\lambda^2\right) f_\omega \left({\bf k}_a,{\bf k}_b\right) - \frac{1}{2 \pi \lambda^2} \int_0^{2 \pi} d \theta \,\xi \left(\lambda^2\right) f_\omega \left({\bf k}_a+\underline{\bf \lambda},{\bf k}_b\right).
\end{eqnarray}
For a sufficiently smooth $f_\omega \left({\bf k}_a,{\bf
k}_b\right)$ this expression is small in the $\lambda
\rightarrow 0$ limit. In fact, the approximation made in
Eq.(7) is only good for a smooth $f_\omega \left({\bf
k}_a,{\bf k}_b\right)$, and the $\lambda$--dependence can ultimately be
studied numerically.
\section{Iterative solution in the NLLA}
\label{Iterating}

The BFKL equation in the NLLA for the forward case as written in
Eq.~(\ref{nll}) can be solved using an iterative procedure similar to the one
applied in \cite{LLAlimit} for the LLA. We will, in the following, show how
this works in the NLLA. 
In the NLLA it becomes meaningful to study the dependence
on the renormalisation scale. Since there are in principle
many different physical scales in the problem, there are
many possible choices also for the renormalisation
scale. Here we will assume that the renormalisation scale is
chosen to depend on the arguments of the Green's function
only
\begin{eqnarray}
\label{mu}
\mu = \mu \left({\bf k}_a^2, {\bf k}_b^2\right).
\end{eqnarray}
Other choices are possible and can be studied in this formalism.
The $\mu$--dependence of the trajectory and the kernel will be explicitly
shown in all of our expressions from now on. It is also convenient to use the
following notation for the kernel
\begin{eqnarray}
\widehat{\mathcal{K}}_r \left({\bf k}_a, {\bf k}_a+{\bf k},\lambda^2,
\mu \left({\bf k}_a^2,{\bf k}_b^2\right)\right) 
&\equiv& \frac{1}{\pi {\bf k}^2} \xi \left({\bf k}^2\right) \theta\left({\bf k}^2-\lambda^2\right)+\widetilde{\mathcal{K}}_r \left({\bf k}_a,{\bf k}_a+{\bf k}\right). 
\end{eqnarray}
Using this notation we can take the $\omega$-dependence to the right hand
side of Eq.~(\ref{nll}), i.e.
\begin{eqnarray}
f_\omega \left({\bf k}_a,{\bf k}_b\right)&=&
\frac{1}{\omega - {\omega}_0 \left({\bf k}_a^2,\lambda^2,\mu 
\left({\bf k}_a^2,{\bf k}_b^2\right)\right)} \left\{\delta^{(2)}\left({\bf k}_a-{\bf k}_b\right) \right.\\
&&\left. \hspace{-2cm}+ \int d^2 {\bf k}~
\widehat{\mathcal{K}}_r \left({\bf k}_a,{\bf k}_a+{\bf k},\lambda^2,\mu\left({\bf k}_a^2,{\bf k}_b^2\right)\right) 
f_\omega \left({\bf k}_a+{\bf k},{\bf k}_b\right) \right\},\nonumber
\end{eqnarray} 
and iterate this expression. This procedure leads to 
\begin{eqnarray}
\label{iterating}
f_{\omega} \left({\bf k}_a ,{\bf k}_b\right) &=& 
\frac{\delta^{(2)} \left({\bf k}_a - {\bf k}_b \right)}{\omega - {\omega}_0 \left({\bf k}_a^2,\lambda^2,\mu \left({\bf k}_a^2,{\bf k}_b^2\right)\right)}\\
&&\hspace{-2cm}+ \int d^2 {\bf k}_1 
\frac{\widehat{\mathcal{K}}_r
\left({\bf k}_a,{\bf k}_a+{\bf k}_1,\lambda^2,\mu \left({\bf k}_a^2
,{\bf k}_b^2\right)\right)}
{\omega - {\omega}_0 \left({\bf k}_a^2,\lambda^2,\mu 
\left({\bf k}_a^2,{\bf k}_b^2\right)\right)} \nonumber\\
&&\hspace{-1cm} \times
\frac{\delta^{(2)} \left({\bf k}_a +{\bf k}_1 - {\bf k}_b\right)}
{\omega - {\omega}_0 \left(\left({\bf k}_a+{\bf k}_1\right)^2,\lambda^2,\mu 
\left(\left({\bf k}_a+{\bf k}_1\right)^2,{\bf k}_b^2\right)\right)} \nonumber\\
&&\hspace{-2cm}+ \int d^2 {\bf k}_1 
\frac{\widehat{\mathcal{K}}_r \left({\bf k}_a,{\bf k}_a+{\bf k}_1,\lambda^2,\mu\left({\bf k}_a^2,{\bf k}_b^2\right)\right)}
{\omega - {\omega}_0 \left({\bf k}_a^2,\lambda^2,\mu \left({\bf k}_a^2,{\bf k}_b^2\right)\right)} \nonumber\\
&&\hspace{-1cm} \times \int d^2 {\bf k}_2 
\frac{\widehat{\mathcal{K}}_r
\left({\bf k}_a + {\bf k}_1,{\bf k}_a+{\bf k}_1+{\bf k}_2,\lambda^2,\mu\left(\left({\bf k}_a+{\bf k}_1\right)^2,{\bf k}_b^2\right)\right)}
{\omega - {\omega}_0 \left(\left({\bf k}_a+{\bf k}_1\right)^2,\lambda^2,\mu\left(\left({\bf k}_a+{\bf k}_1\right)^2,{\bf k}_b^2\right)\right)} \nonumber\\
&& \hspace{-1cm}\times
\frac{\delta^{(2)} \left({\bf k}_a +{\bf k}_1+{\bf k}_2 - {\bf k}_b\right)}
{\omega - {\omega}_0 \left(\left({\bf k}_a+{\bf k}_1+{\bf k}_2\right)^2,\lambda^2,\mu\left(\left({\bf k}_a+{\bf k}_1+{\bf k}_2\right)^2,{\bf k}_b^2\right)\right)} \nonumber\\
&&\hspace{-2cm}+ \cdots \nonumber
\end{eqnarray}

As we want to obtain the final solution in energy space we perform the
inverse Mellin transform of Eq.~(\ref{iterating}) as defined in
Eq.~(\ref{Mellin}). In this way it is possible to obtain the following
expression for the solution in the NLLA\footnote{Using the notation $y_0
  \equiv \Delta$.}:
\begin{eqnarray}
\label{ours}
f({\bf k}_a ,{\bf k}_b, \Delta) 
&=& \exp{\left(\omega_0 \left({\bf k}_a^2,{\lambda^2},\mu\left({\bf k}_a^2,{\bf k}_b^2\right)\right) \Delta \right)}
\left\{\frac{}{}\delta^{(2)} ({\bf k}_a - {\bf k}_b) \right. \\
&&\hspace{-3cm}+ \sum_{n=1}^{\infty} \prod_{i=1}^{n} 
\int d^2 {\bf k}_i \left[\frac{\theta\left({\bf k}_i^2 - \lambda^2\right)}{\pi {\bf k}_i^2} \xi\left({\bf k}_i^2,\mu\left(\left({\bf k}_a+\sum_{l=0}^{i-1}{\bf k}_l\right)^2,{\bf k}_b^2\right)\right) \right. \nonumber\\
&&\hspace{1cm}+\left. \widetilde{\mathcal{K}}_r \left({\bf k}_a+\sum_{l=0}^{i-1}{\bf k}_l,
{\bf k}_a+\sum_{l=1}^{i}{\bf k}_l,\mu \left(\left({\bf k}_a+\sum_{l=0}^{i-1}{\bf k}_l\right)^2,{\bf k}_b^2\right)\right)\frac{}{}\right]\nonumber\\
&& \hspace{-3cm} \times  
\int_0^{y_{i-1}} d y_i ~ {\rm exp}\left[\left(
\omega_0\left(\left({\bf k}_a+\sum_{l=1}^i {\bf k}_l\right)^2,\lambda^2,
\mu \left(\left({\bf k}_a+\sum_{l=0}^{i}{\bf k}_l\right)^2,{\bf k}_b^2\right)\right)\right.\right.\nonumber\\
&&\left.\left.\left.\hspace{-3cm}
-\omega_0\left(\left({\bf k}_a+\sum_{l=1}^{i-1} {\bf k}_l\right)^2,
{\lambda^2},\mu\left(\left({\bf k}_a+\sum_{l=0}^{i-1}{\bf k}_l\right)^2,{\bf k}_b^2\right)\right)\right) y_i\right] \delta^{(2)} \left(\sum_{l=1}^{n}{\bf k}_l 
+ {\bf k}_a - {\bf k}_b \right)\right\}. \nonumber
\end{eqnarray}

The expression (\ref{ours}) is only weakly $\lambda$--dependent in the
$\lambda \to 0$ limit. We can show this point by applying it to a given
smooth function, $\Phi \left({\bf k}_a,{\bf k}_b \right)$, and selecting
those terms proportional to $\Delta$, i.e.
\begin{eqnarray}
\int d^2{\bf k}_b ~ f\left({\bf k}_a,{\bf k}_b,\Delta\right) 
\Phi \left({\bf k}_a,{\bf k}_b \right) &=&
\Phi \left({\bf k}_a, {\bf k}_a\right) + \Delta  \left\{ \frac{}{}\omega_0 \left({\bf k}_a^2,\lambda^2\right) 
\Phi \left({\bf k}_a, {\bf k}_a\right) \right. \\
&&\hspace{-5cm}\left. + \int d^2{\bf k}_1 
\left(\frac{\theta\left({\bf k}_1^2 - \lambda^2\right)}{\pi {\bf k}_1^2} 
\xi \left({\bf k}_1^2\right) + \widetilde{\mathcal{K}}_r 
\left({\bf k}_a,{\bf k}_a+{\bf k}_1\right)\right) 
\Phi \left({\bf k}_a, {\bf k}_a+{\bf k}_1\right)\right\} 
+ \cdots \nonumber
\end{eqnarray}
The $\lambda$--dependence of this expression can be studied
in a similar manner to the one applied for Eq.(18).

As a final remark we would like to point out that our solution in the LLA
limit is consistent with similar results in the LLA in the literature
\cite{LLAlimit}. If we take the leading--logarithmic limit of our expressions
we have
\begin{eqnarray}
\omega_0({\bf q}^2,{\lambda^2}) ~=~ 
- \bar{\alpha}_s \ln{\frac{{\bf q}^2}{{\lambda^2}}}, ~~~ 
\xi ~=~ \bar{\alpha}_s, ~~~\eta ~=~ 0, ~~~ \widetilde{\mathcal{K}}_r  
\left({\bf q},{\bf q}'\right) ~=~ 0.
\end{eqnarray}
In this case the solution takes the simple form
\begin{eqnarray}
f({\bf k}_a ,{\bf k}_b, \Delta) 
&=& \left(
\frac{\lambda^2}{k_a^2}\right)^{\bar{\alpha}_s \,\Delta}
\left\{\frac{}{}\delta^{(2)} ({\bf k}_a - {\bf k}_b) \right. \\
&&\hspace{-3.4cm}+ \sum_{n=1}^{\infty} \prod_{i=1}^{n} 
\bar{\alpha}_s \int d^2{\bf k}_i 
\frac{\theta\left({\bf k}_i^2- \lambda^2\right)}{\pi {\bf k}_i^2} 
\int_0^{y_{i-1}} d y_i 
\left(\frac{\left({\bf k}_a+\sum_{l=1}^{i-1} {\bf k}_l\right)^2}
{\left({\bf k}_a+\sum_{l=1}^i {\bf k}_l\right)^2}\right)
^{\bar{\alpha}_s \, y_i} \nonumber\\
&&\left. \times \, \delta^{(2)} \left(\sum_{l=1}^{n}{\bf k}_l 
+ {\bf k}_a - {\bf k}_b \right)  \right\}.\nonumber
\end{eqnarray}
This expression is equivalent to that obtained in \cite{LLAlimit} in the LLA
where the regularisation of the infrared divergences was performed in a
different manner. The main result of this Letter is the extension of this
iterative method to the NLLA. This has been possible by using the NLLA BFKL
equation in $4+2 \epsilon$ dimensions to introduce a cut--off in the phase
space. This renders the $\epsilon \to 0$ limit finite and thus allows us to
iterate the BFKL equation to obtain the solution in Eq.~(\ref{ours}).

\section{Conclusions}
\label{conclusions}

We have presented a procedure to solve the BFKL equation for forward
scattering in the next--to--leading logarithmic approximation. We have shown
how it is important to use the kernel in dimensional regularisation and
introduce a cut--off, $\lambda$, in the phase space, Eq.~(\ref{nll}). This
allows us to write the solution in a compact form, Eq.~(\ref{ours}), suitable
for numerical studies, which will be presented in a future work. We have also
shown the mechanism by which the solution is weakly dependent on the cut--off
for small values of $\lambda$. We would like to point out that we keep the
full angular information in our solution by solving the equation for any
conformal spin, i.e. we do not use the angular--averaged kernel (see
Eq.~(\ref{non_ang_av})). This will allow the study of spin--dependent
observables in the NLLA.

Work is in progress to understand the BFKL resummed gluon Green's function
using this approach, and to quantify the effect of those terms related to the
running of the coupling \cite{running} compared to the scale invariant ones.
A study of the solution in the NLLA of the N=4 supersymmetric
case~\cite{SUSY}, where the coupling does not run, is also under
consideration.

The ultimate goal in the application of our solution is the calculation of
cross--sections in the NLLA for those processes where the BFKL resummation
should be relevant. The presented solution has the advantage of not operating
in the Mellin space, i.e. we can use the NLLA impact factors
\cite{impactfactors} directly in transverse momentum space in order to make
phenomenological predictions. It will, in principle, be possible to
disentangle the structure of the final state allowing the study of, e.g.,
multiplicities, extending the work of~\cite{phenomenology} to the NLLA. As a
final point we would like to indicate that this approach is also suitable for
studying the solution of the BFKL equation in the NLLA for the non-forward
case when the kernel for this case has been completed.

\noindent {\bf Acknowledgements}
We would like to thank Victor Fadin for very useful correspondence and Stefan
Gieseke for discussions. ASV wishes to thank Jochen Bartels and Lev Lipatov
for inspiring discussions and the II. Institut f{\" u}r Theoretische Physik
at the University of Hamburg and the CERN Theory Division for hospitality,
and acknowledges the support of PPARC (Posdoctoral Fellowship:
PPA/P/S/1999/00446).

\end{document}